 \documentclass[final,3p,times,twocolumn]{elsarticle}

 \usepackage{graphics}

 \usepackage{xcolor}
\usepackage{amssymb}
\usepackage{amsmath}
\usepackage{cases}
\usepackage{epsfig,subfigure}
\usepackage{graphicx}
\usepackage{epstopdf}

 \biboptions{comma,square,compress,numbers,sort}

\journal{Physics Letters B}

\begin{document}

\begin{frontmatter}



\title{Thick branes with inner structure in mimetic gravity}


\author[a]{Yi Zhong}
\ead{zhongy13@lzu.edu.cn}
\author[b]{Yuan Zhong}
\ead{zhongy@mail.xjtu.edu.cn}
\author[a]{Yu-Peng Zhang}
\ead{zhangyupeng14@lzu.edu.cn}
\author[a]{Yu-Xiao Liu\corref{cor4}}
  \ead{liuyx@lzu.edu.cn}
  \cortext[cor4]{The corresponding author.}

\address[a]{Institute of Theoretical Physics, Lanzhou University, Lanzhou 730000, P. R. China}
\address[b]{School of Science, Xi'an Jiaotong University, Xi'an 710049, P. R. China}

\begin{abstract}
In this paper, thick branes generated by mimetic scalar field are investigated. Three typical thick brane models are constructed and the linear tensor and scalar perturbations are analyzed. These branes have different inner structures, some of which are absent in general relativity. For each brane model, the solution is stable under both tensor and scalar perturbations. The tensor zero modes are localized on the branes, while the scalar perturbations do not propagate and they are not localized on the brane. As the branes split into multi sub-branes for specific parameters, the potentials of the tensor perturbations also split into multi-wells, and this may lead to new phenomenon in the resonance of the tensor perturbation and the localization of matter fields.
\end{abstract}


\end{frontmatter}


\section{Introduction}
Though the standard model of cosmology is in good agreement with observation
data and has made a number of successful predictions, it still faces severe problems. In this model, dark matter constitutes $84.5 \%$ of total mass of matter. However, dark matter has never been directly observed and its nature remains unknown. One possible explanation for dark matter is that Einstein's gravity is modified at large scale. Among the modified theories of gravity, mimetic gravity is a particularly interesting one and has been investigated widely. In mimetic gravity, the
physical metric $g_{\mu\nu}$ is defined in terms of an auxiliary metric $\hat{g}_{\mu\nu}$
and a scalar field $\phi$ by
$g_{\mu\nu}=-\hat{g}_{\mu\nu}\hat{g}^{\alpha\beta}\partial_\alpha\phi\partial_\beta\phi$ \cite{Chamseddine:2013kea}.
By this means, the conformal degree of freedom is separated in a covariant way, and
this extra degree of freedom becomes dynamic and can mimic cold dark matter \cite{Chamseddine:2013kea,Chamseddine:2014vna}. Furthermore, it is possible to unify the late-time acceleration and inflation within this framework \cite{Nojiri_2014,Momeni:2015gka,Odintsov:2015cwa,Odintsov:2015ocy}. To obtain a viable theory confronted with the cosmic evolution, this theory is transformed to Lagrange multiplier formulation and the potential of the mimetic scalar field is considered. For more recent works of mimetic gravity see Refs. \cite{Momeni:2015gka,Matsumoto:2015wja,Odintsov:2015ocy,Nojiri:2016ppu,Odintsov:2015wwp,Oikonomou:2015lgy,Cognola:2016gjy,Rabochaya_2016,Odintsov:2016imq} or Ref. \cite{Sebastiani:2016ras} for a review.

On the other hand, the brane world scenario has been an attractive topic in the last two decades, since the Randall-Sundrum (RS) model being proposed \cite{Randall:1999ee,Randall:1999vf}. It is shown that the
gauge hierarchy problem and the cosmological constant problem can be explained in  this model \cite{Randall:1999ee,Randall:1999vf,Kim:2000mc}. Various extensions of the RS model have been investigated in Refs. \cite{Davoudiasl:1999tf,Shiromizu:1999wj,Gherghetta:2000qt,Rizzo:2010zf,Yang:2012dd,Agashe:2014jca}. In these models, the brane is considered to be geometrically thin. However, we cannot consider a geometrically thin brane, as it is believed that there exists a minimum length scale. For this reason, thick brane models were proposed \cite{Csaki:2000fc,DeWolfe:1999cp,Gremm:1999pj} and investigated thoroughly \cite{Liu:2017gcn}.

Recently, Sadeghnezhad and Nozari investigated the late time cosmic expansion and inflation on a thin brane in mimetic gravity \cite{Sadeghnezhad:2017hmr}. It is necessary to investigate thick brane in this theory. In the thick brane world scenario, the brane can be a domain wall generated by a background scalar field \cite{Csaki:2000fc,DeWolfe:1999cp,Gremm:1999pj,Afonso:2006gi,Afonso:2007zz,Guo:2014bxa,German:2013sk,Liu:2009ega,Dzhunushaliev:2009va} or by pure geometry in a co-dimension one space-time \cite{Arias:2002ew,BarbosaCendejas:2006hj,Liu:2011am,Dzhunushaliev:2009dt,Zhong:2015pta}. Thus, it is natural to generate domain wall by the mimetic scalar field, and the new degree of freedoms allows us to construct new type of thick branes. For this reason, we will investigate several thick branes in mimetic gravity and examine stability under tensor and scalar perturbations. We will find that some of the thick branes have very different inner structures from the case of general relativity.

The organization of this paper is as follows. In Sec.~\ref{Sec2}, we construct three flat thick brane models. In Sec.~\ref{Sec3} we consider the behavior of the tensor perturbations in each of the brane models. In Sec.~\ref{Sec4} we analyze the scalar perturbations. Finally, the conclusion and discussion are given in Sec.~\ref{SecConclusion}.

\section{Construction of the thick brane models} \label{Sec2}

In the natural unit, the action of the mimetic gravity is
 \begin{eqnarray}
        S\!=\!\int d^4x dy\sqrt{-g}\left( \frac{R}{2}
        + L_{\phi} \right),
        \label{action mgb1}
    \end{eqnarray}
where the lagrangian of the mimetic scalar field is
\begin{eqnarray}
L_{\phi}=\lambda\left[g^{MN}\partial_M \phi \partial_N \phi-U(\phi)\right]-V(\phi),        \label{Lagrangian}
\end{eqnarray}
and the $\lambda$ is a Lagrange multiplier. In the original mimetic gravity, $U(\phi)=-1$ \cite{Chamseddine:2013kea}, and then it is extended into to the case with $U(\phi)<0$ \cite{Astashenok2015}. In thick brane models, a brane will be generated by the mimetic scalar field $\phi=\phi(y)$. Therefore, we assume that $U(\phi)=g^{MN}\partial_M \phi \partial_N \phi>0$.
The equations of motion (EoM) are obtained by varying the above action with respect to $g_{MN}$, $\phi$ and $\lambda$ respectively:
    \begin{eqnarray}
        \label{var eom1}
        G_{MN}+2\lambda \partial_M \phi \partial_N \phi-L_{\phi}g_{MN}=0,    \\
        \label{var eom2}
        2\lambda\Box^{(5)}\phi+2\nabla_{M}\lambda\nabla^{M}\phi+\lambda \frac{\partial U}{\partial \phi}+\frac{\partial V}{\partial \phi}=0, \\
        \label{var eom3}
        g^{MN}\partial_M \phi \partial_N \phi-U(\phi)=0.
    \end{eqnarray}
Here the five-dimensional d'Alembert operator is defined as  $\Box^{(5)}=g^{MN}\nabla_{M}\nabla_{N}$. The indices $M,N\cdots=0,1,2,3,5$ denote the bulk coordinates and $\mu,\nu\cdots$ denote the ones on the brane.

In this paper we consider the following brane world metric which preserves $4$-dimensional Poincare invariance
    \begin{eqnarray}
        \label{brane metric1}
       ds^2=a^2(y)\eta_{\mu\nu}dx^{\mu}dx^{\nu}+dy^2.
    \end{eqnarray}
With this metric assumption, Eqs. (\ref{var eom1})-(\ref{var eom3}) read
    \begin{eqnarray}
        \label{eom21}
      \frac{3a'^2}{a^2}+\frac{3a''}{a}+V(\phi)+\lambda \left(U(\phi)-\phi'^2\right)=0, \\
       \label{eom22}
       \frac{6a'^2}{a^2}+V(\phi)+2\lambda \left(U(\phi)+\phi'^2\right)=0, \\
       \label{eom23}
       \lambda \left(\frac{8a'\phi'}{a}+2\phi''+\frac{\partial U}{\partial \phi}\right)
        +2\lambda'\phi'+\frac{\partial V}{\partial \phi}=0,  \\
       \label{eom24}
       \phi'^2=U(\phi).
    \end{eqnarray}
Here, the primes denote the derivatives with respect to the extra dimension coordinate $y$.
Substituting Eqs. (\ref{eom21}) and (\ref{eom24}) into Eq. (\ref{eom22}) we can solve the Lagrange multiplier $\lambda(y)$
    \begin{eqnarray}
        \label{sol lambda}
        \lambda = \frac{3(-a'^2+aa'')}{2a^2\phi'^2}.
    \end{eqnarray}
Note that there are only three independent equations in Eqs. (\ref{eom21})-(\ref{eom24}). Once $A(y)$ and $\phi(y)$ are given, we can get $\lambda(y)$, $V(\phi)$ and $U(\phi)$ from Eqs. (\ref{sol lambda}), (\ref{eom21}) and (\ref{eom24}) respectively. Next, we will investigate three kinds of thick brane models.

\subsection{Model 1}

\begin{figure}[!htb]
    \begin{center}
    \subfigure[The warp factor]{
        \includegraphics[width=3.6cm]{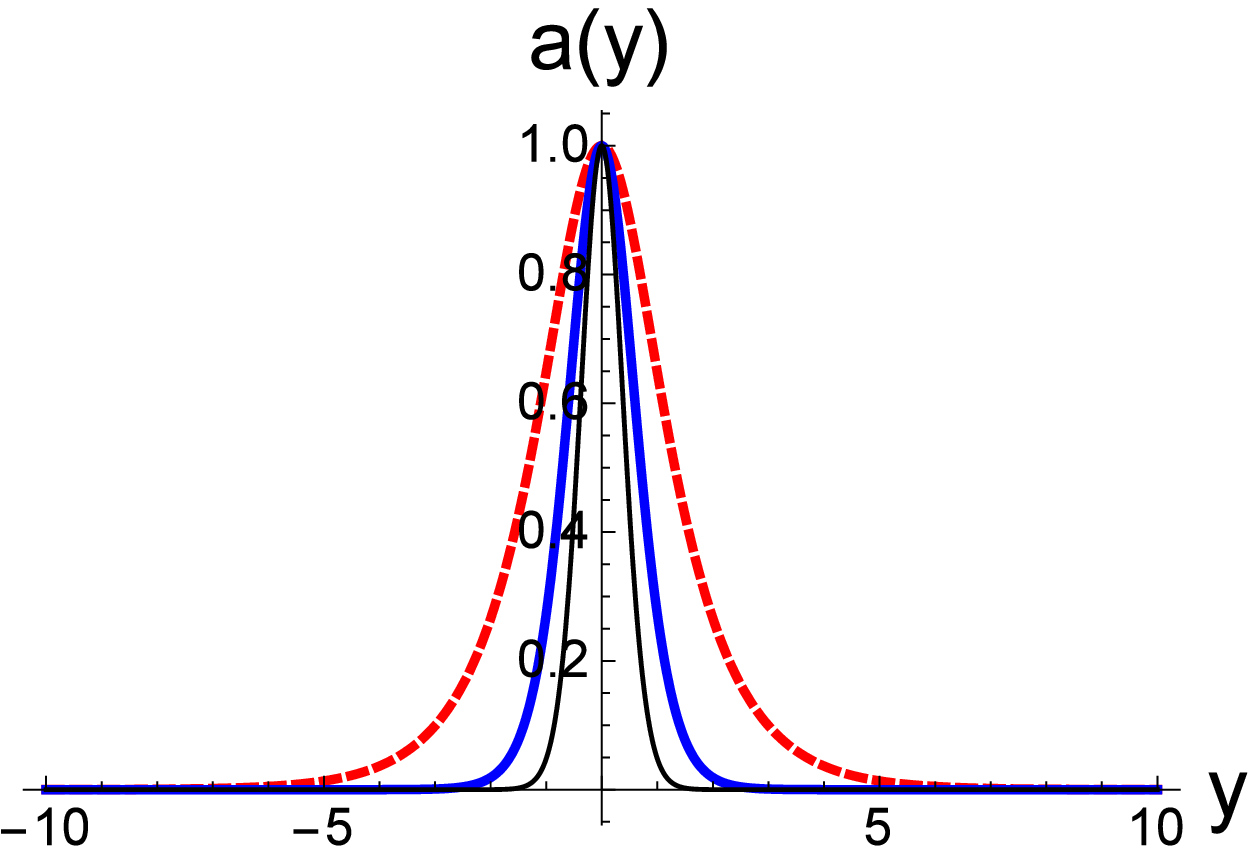}}
    \subfigure[The  scalar field]{
        \includegraphics[width=3.6cm]{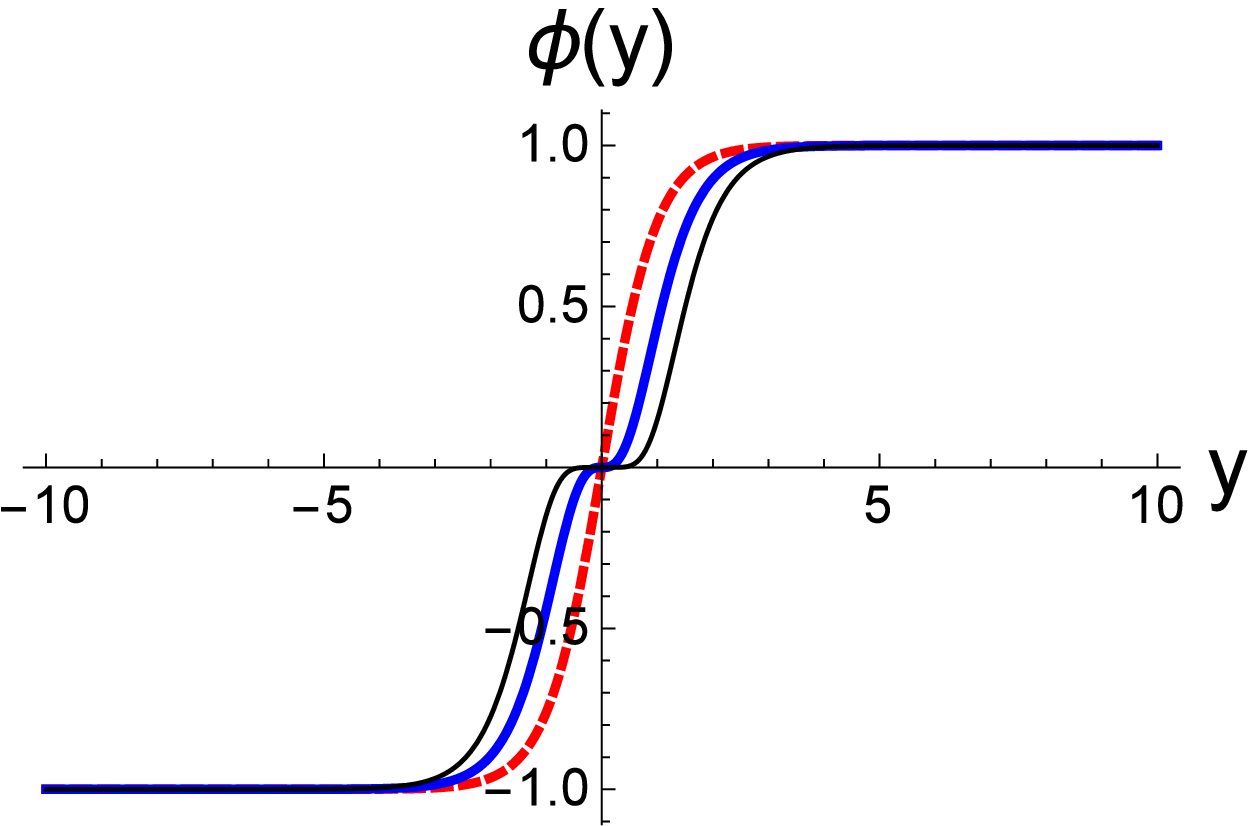}}
    \end{center}
    \caption{The shapes of the warp factor $a(y)$ and the scalar field $\phi(y)$ of the first brane model. The parameters are set as $k=1$, $v=1$
             and $n=1$ for the dashed red lines, $n = 3$ for the thick
             blue lines, and $n=7$ for the thin black lines.}\label{One}
    \end{figure}

In the first model, we consider the solution of the warp factor $a(y)$ and the scalar field that has similar property as the case of general relativity. The solution of such model is given by
    \begin{eqnarray}
        a(y)\!\!&=\!\!& \text{sech}^n(ky), \\
        \phi(y)\!\!&=\!\!& v\,\text{tanh}^n(ky), \\
        \lambda(y)\!\!&=\!\!& -\frac{3}{2nv^2} \text{sinh}^2(ky) \text{tanh}^{-2n}(ky), \\
        V(\phi)\!\!&=\!\!& 3k^2\left[ n-n(1+2n)\Big(\frac{\phi}{v}\Big)^{\frac{2}{n}}\right], \\
        U(\phi)\!\!&=\!\!& k^2 n^2 v^2 \left(\frac{\phi}{v}\right)^{\frac{2(n-1)}{n}}
        \left[\left(\frac{\phi}{v}\right)^{\frac{2}{n}}-1\right]^2,
    \end{eqnarray}
where $n$ is a positive odd integer. The shapes of the warp factor $a(y)$ and the scalar field $\phi(y)$ are plotted in Fig.~\ref{One}, from which we can see that the double-kink scalar field $\phi$ generates a single brane.

\subsection{Model 2}

Next we would like to construct a model with multi sub-branes, for which the warp factor has many maxima while the scalar field is still a single kink. One typical solution of such model is given by
    \begin{eqnarray}
        \!\!\!\!a(y)\!\!&=\!\!& \text{sech}(k(y-b))\nonumber\\
        &&+\text{sech}(ky)+\text{sech}(k(y-b)), \\
        \!\!\!\!\phi(y)\!\!&=\!\!& v~\text{tanh}(ky), \\
        \!\!\!\!U(\phi)\!\!&=\!\!& \frac{k^2}{v^2}(\phi^2-v^2)^2.
    \end{eqnarray}
Here we do not show the complicate expressions of $\lambda(y)$ and $V(\phi)$. Note that $\lambda(y)$ can be solved from Eq.~\eqref{sol lambda}, and $V(\phi)$ is given by $V(\phi(y)) = -\frac{6a'^2}{a^2} - 4\lambda U(\phi)$ with the replacement $y\rightarrow \frac{1}{k}{\tanh ^{-1}\left({\phi }/{v}\right)}$.
The shape of the warp factor of this model is shown in Fig. \ref{fig model 21}, from which it can be seen that small parameter $b$ corresponds to a single brane and the brane will split into three sub-branes as the parameter $b$ increases. The distance between two sub-branes is $b$.

Furthermore, this model can be extended to a brane array described by the following warp factor
    \begin{eqnarray}
        a(y)\!\!\!\!&=\!\!\!\!& \sum_{n=-N}^{N} \text{sech}(k(y+nb)),
    \end{eqnarray}
where $N$ is an arbitrary positive integer. Note that the above solution corresponds to the case of odd number of sub-branes. It is not difficult to construct solution for the case of even number. In addition, we only consider the simple solution for which each part of the warp factor has the same maximum.

\subsection{Model 3}
\begin{figure}[!htb]
    \begin{center}
    \subfigure[model 2]{\label{fig model 21}
        \includegraphics[width=3.6cm]{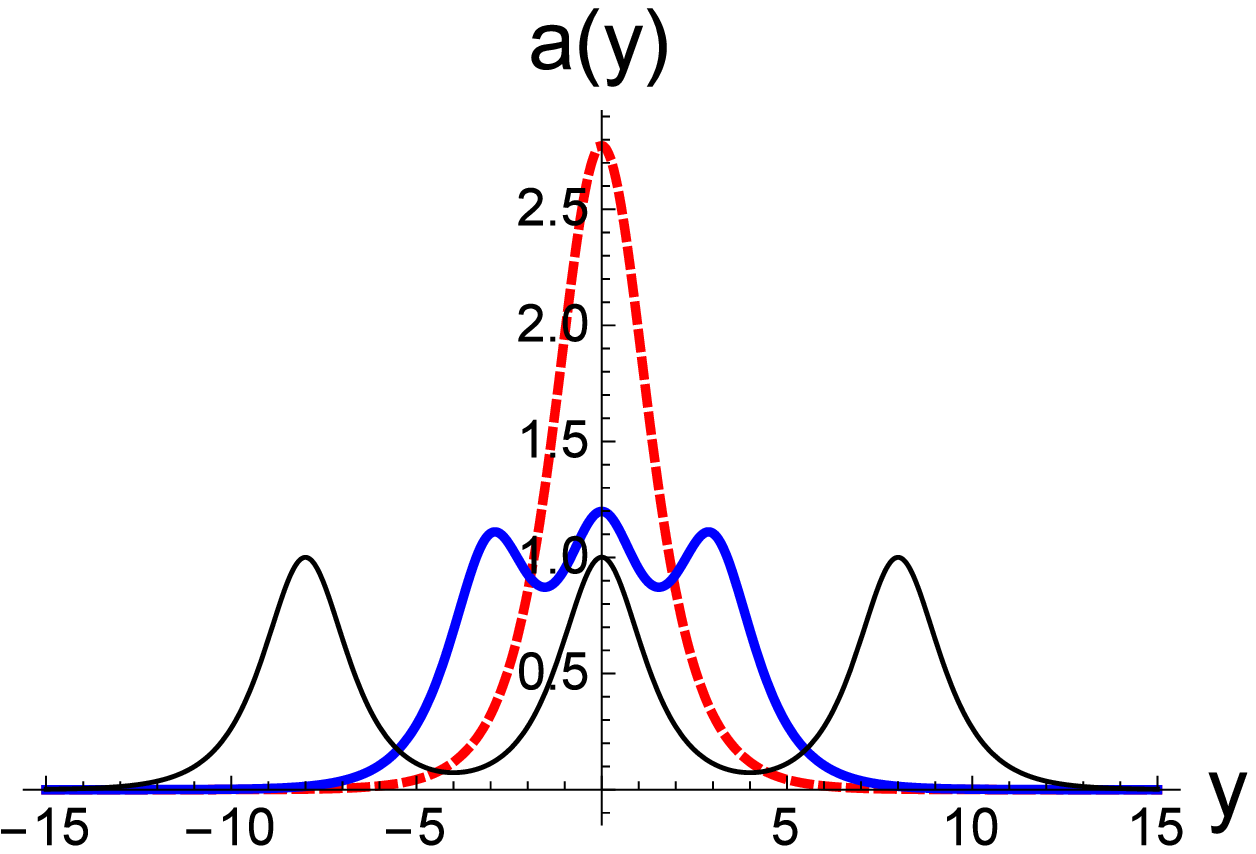}}
    \subfigure[model 3]{\label{fig model 31}
        \includegraphics[width=3.6cm]{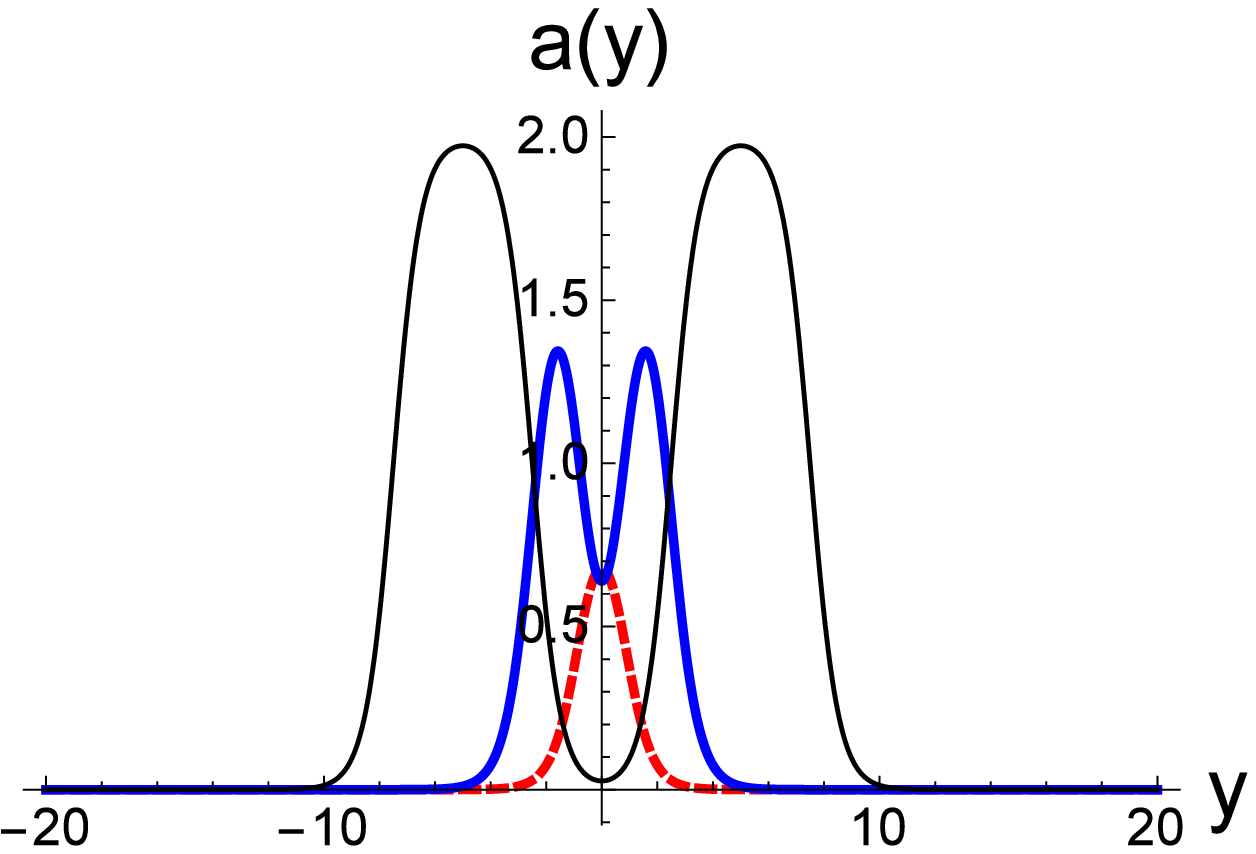}}
    \end{center}
    \caption{The shape of the warp factor $a(y)$  of the brane models 2 and  3. In Fig. (a), the parameters are set as $k=1$, and $b=0.5$ for the dashed red line, $b=3$ for the thick
             blue line, $b=8$ for the thin black line. In Fig. (b), the parameters are set as $k=1$,
             and $b=0.2$ for the dashed red line, $b=0.8$ for the thick
             blue line, $b=2.5$ for the thin black line.}
    \end{figure}

Finally we try to construct another kind of brane solution that will result in different effective potential for the tensor perturbation from the previous model (see the next section). In such model, there is an inner structure in the effective potential for each sub-brane. One typical solution of such brane model with double kink scalar is given by
    \begin{eqnarray}
        a(y)\!\!&=\!\!& \text{tanh}[k(y+3b)]-\text{tanh}[k(y-3b)] \nonumber \\
              \!\!\!\!&&  - \text{tanh}[k(y+b)]+\text{tanh}[k(y-b)], \\
                \phi(y)\!\!&=\!\!& v\,\text{tanh}^n(ky), \\
        U(\phi)\!\!&=\!\!& k^2 n^2 v^2 \left(\frac{\phi}{v}\right)^{\frac{2(n-1)}{n}}
        \left[\left(\frac{\phi}{v}\right)^{\frac{2}{n}}-1\right]^2.
    \end{eqnarray}
Here we do not show the complicate expressions of $\lambda(y)$ and $V(\phi)$.
The shape of the warp factor of this model is shown in Fig. \ref{fig model 31}. The distance of the two sub-branes (for large $b$) is about $6b$ and the width of each sub-brane is $b$. Note that the sub-brane here is fatter than the one in the second model, which results in different structures of the effective potential for each sub-brane in the two models.

Furthermore, this model can be extended into a brane array described by the warp factor
    \begin{eqnarray}
        a(y)\!\!\!\!&=&\!\!\!\! \sum_{n=-N-1}^N \text{tanh}\left[k(y+(2n+1)b)\right],
    \end{eqnarray}
where $N$ is an arbitrary integer.

\section{Tensor perturbation} \label{Sec3}
\begin{figure}[!htb]
    \begin{center}
    \subfigure[~$n=1$]{\label{figure tensor11}
        \includegraphics[width=3.6cm]{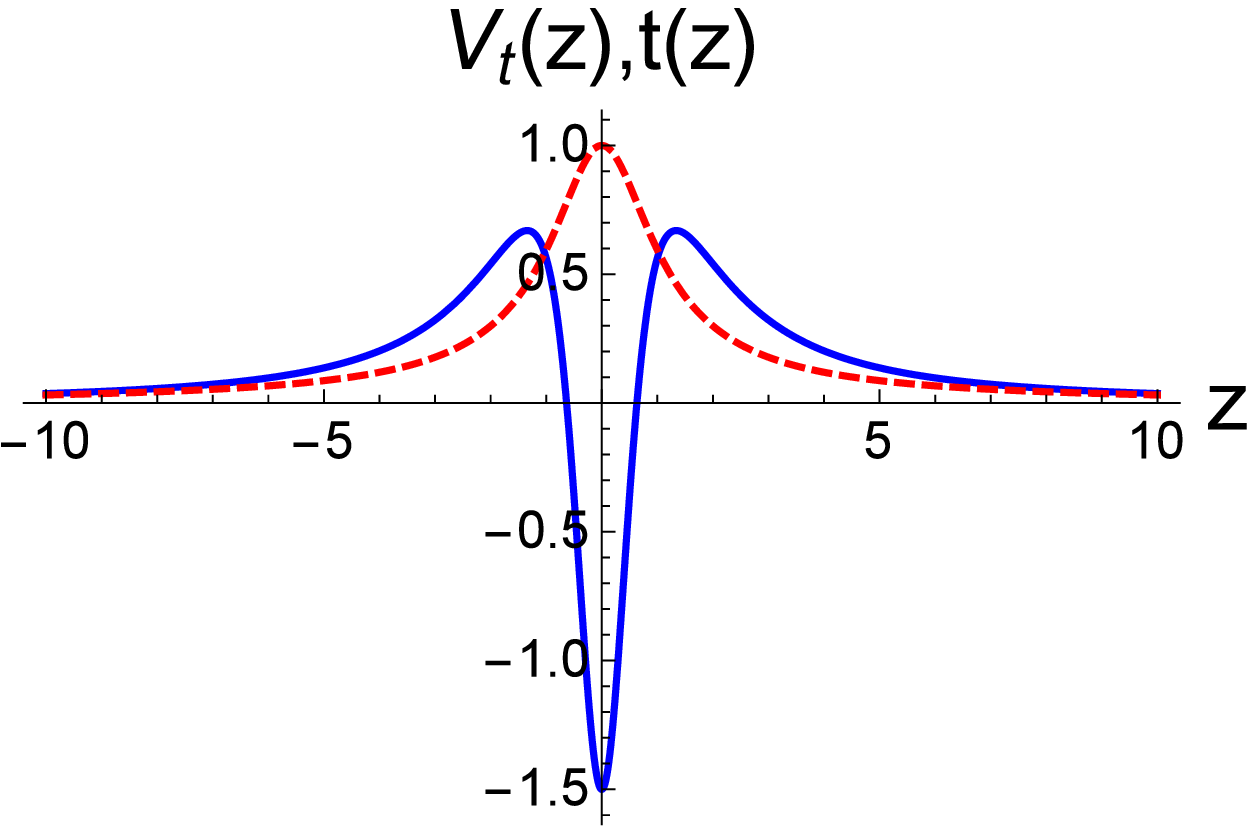}}
    \subfigure[~$n=3$]{\label{figure tensor12}
        \includegraphics[width=3.6cm]{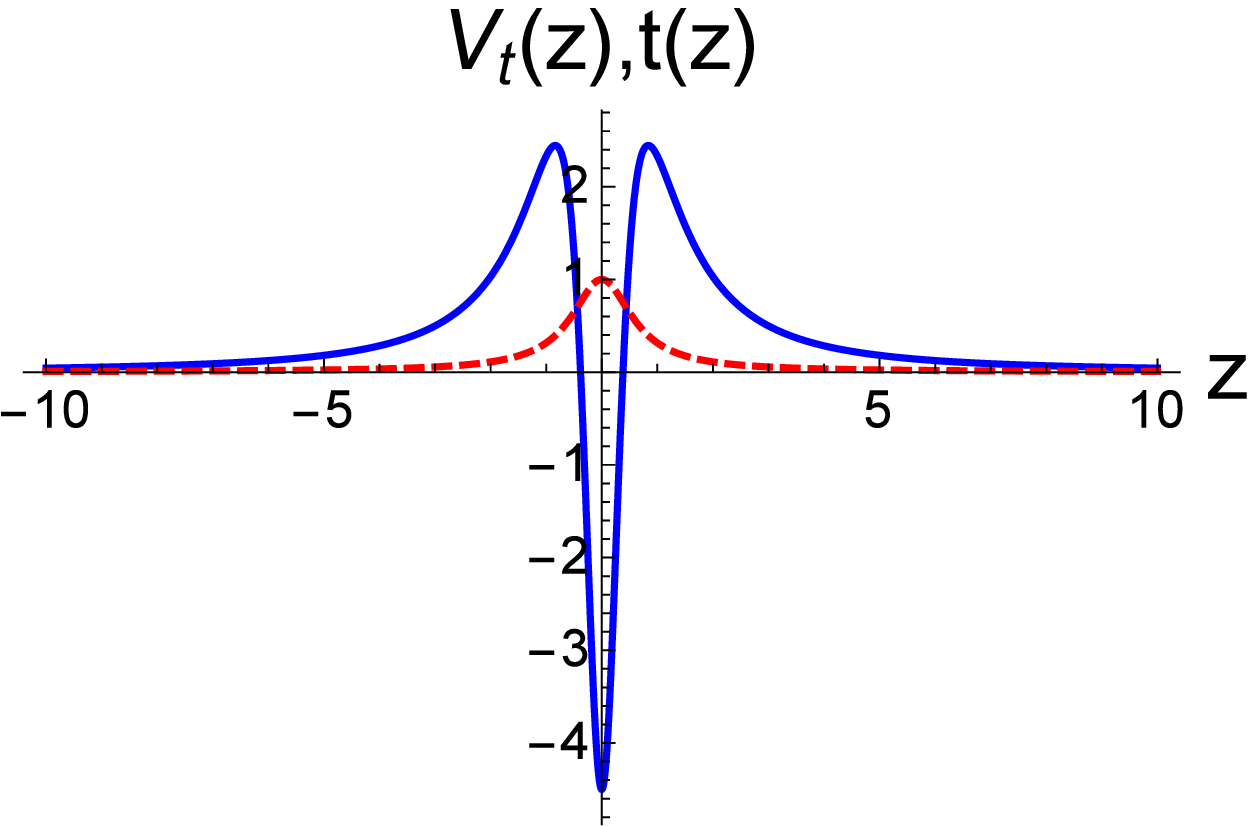}}
    \subfigure[~$n=5$]{\label{figure tensor13}
        \includegraphics[width=3.6cm]{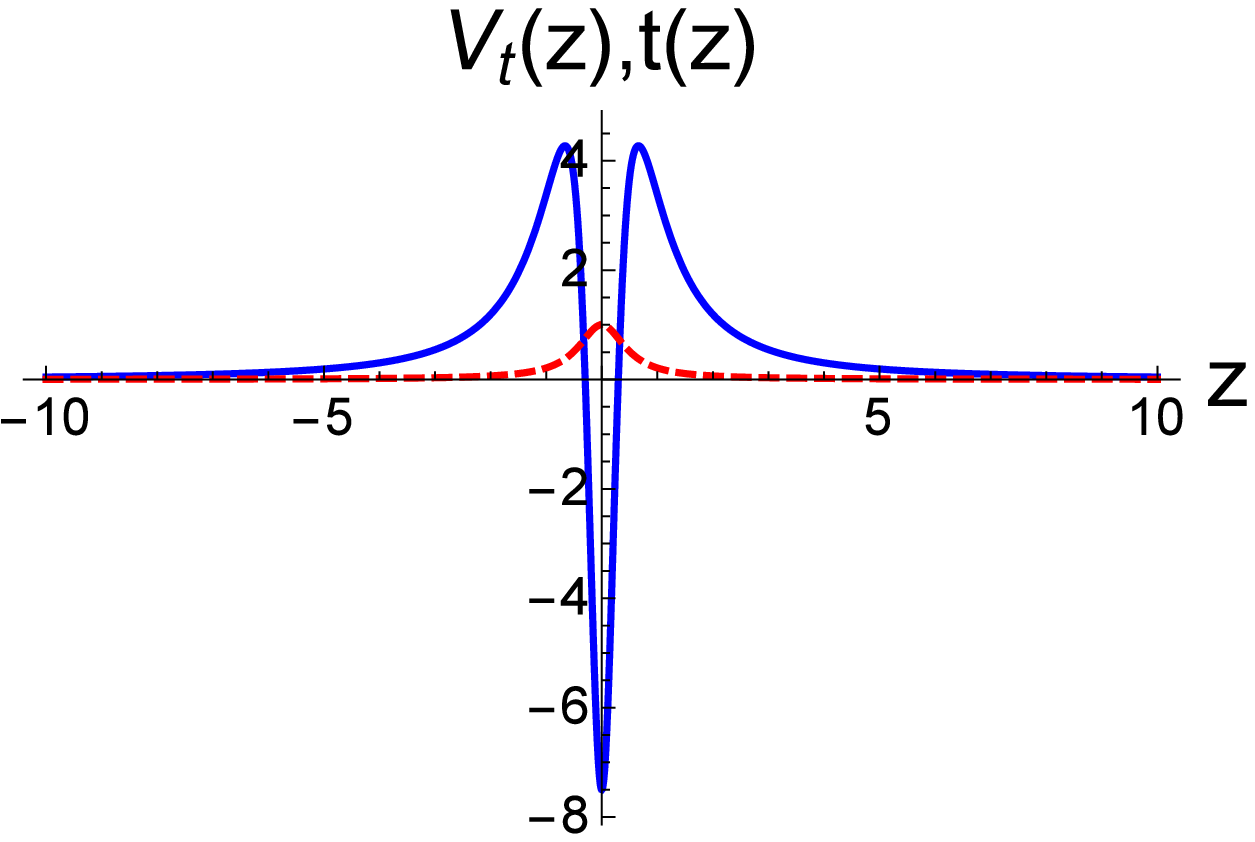}}
    \end{center}
    \caption{The effective potential  $V_t(z)$ (blue solid lines) and the zero mode $t_0(z)$ (red dashed lines) of the tensor perturbation for brane model 1. The parameters are set as $k=1$,
             and $n=1$ in Fig. (a), $n=3$ in Fig. (b), $n=5$ in Fig. (c).} \label{tensorThree}
    \end{figure}

\begin{figure}[!htb]
    \begin{center}
    \subfigure[~$b=0.5$]{\label{figure tensor21}
        \includegraphics[width=3.6cm]{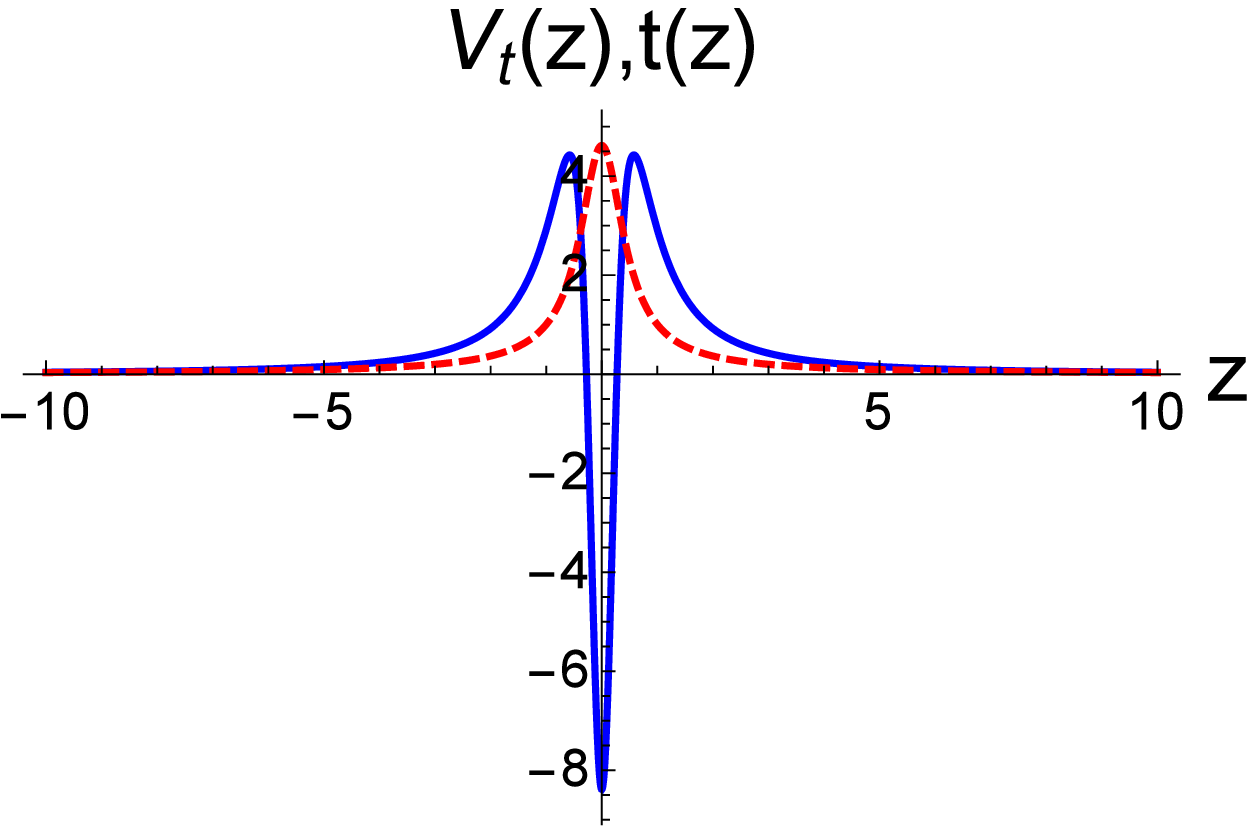}}
    \subfigure[~$b=3$]{\label{figure tensor22}
        \includegraphics[width=3.6cm]{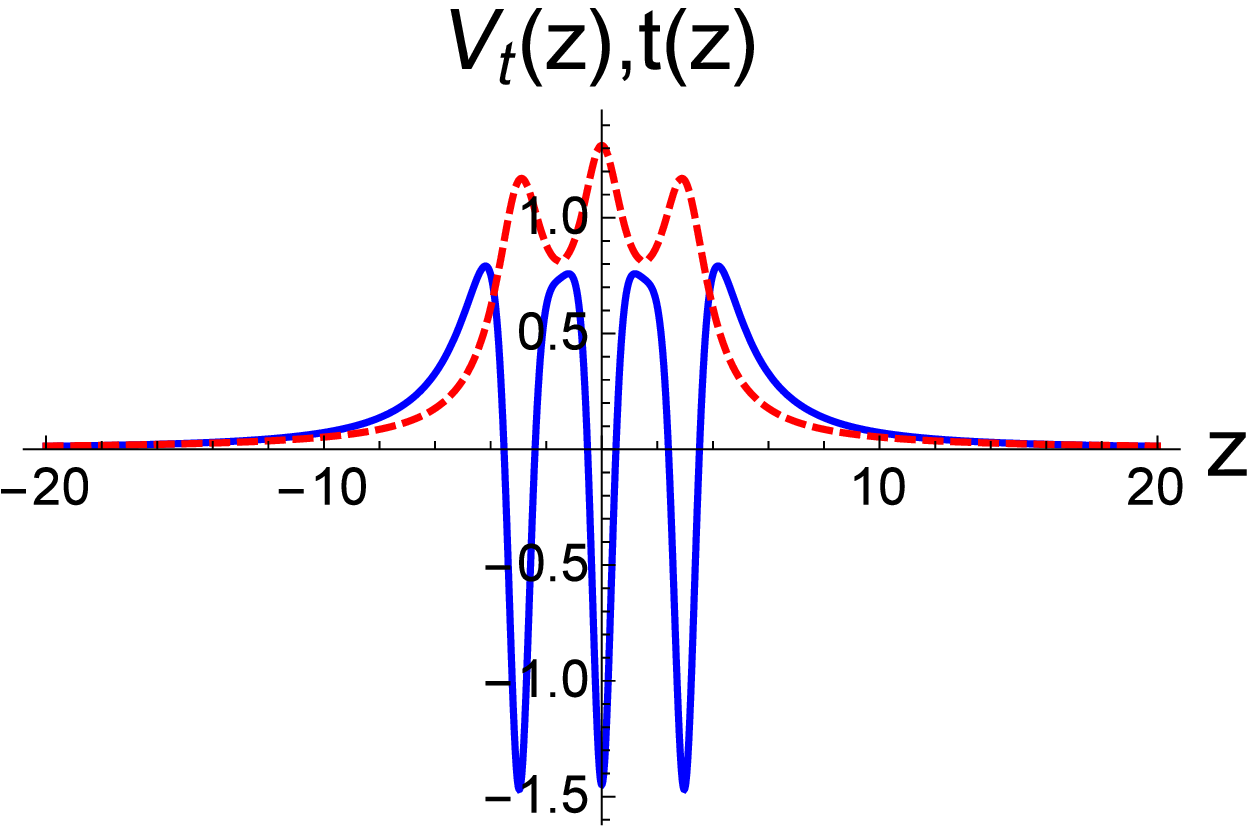}}
    \subfigure[~$b=8$]{\label{figure tensor23}
        \includegraphics[width=3.6cm]{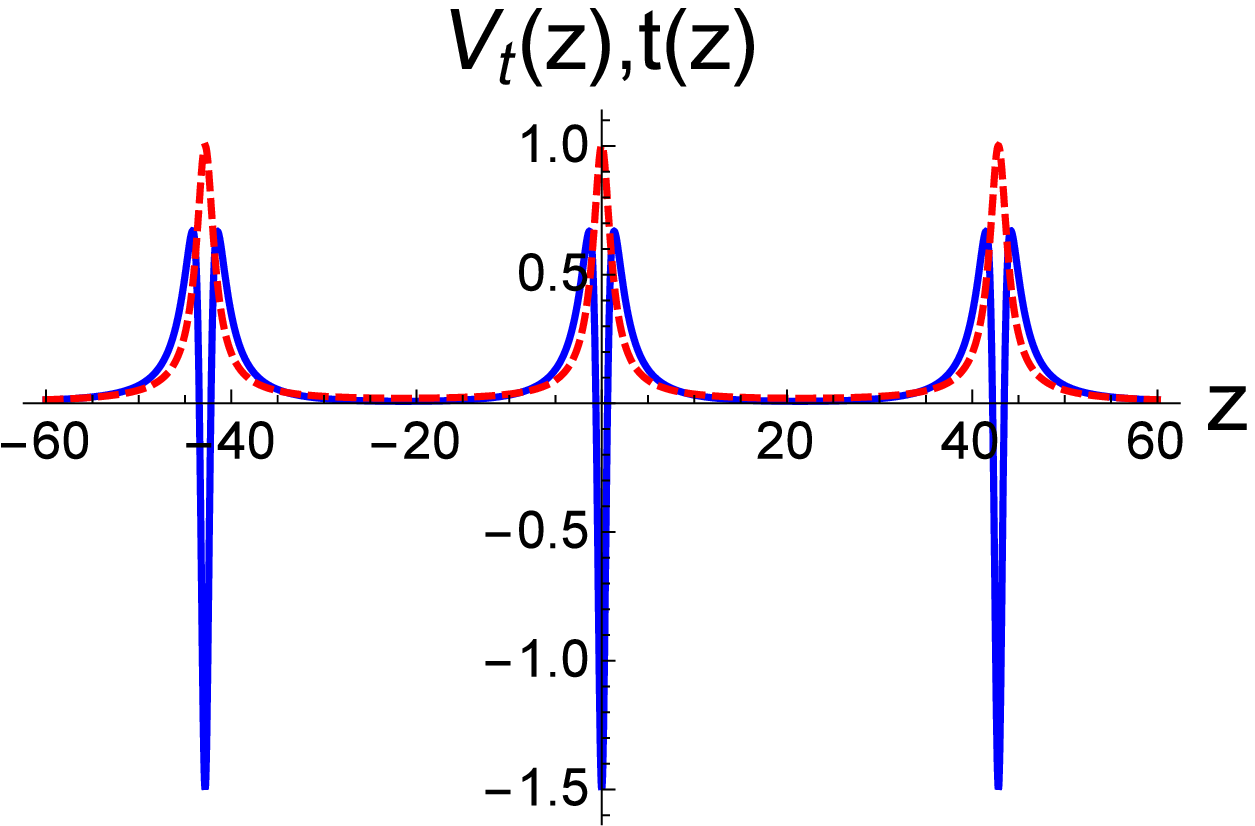}}
    \end{center}
    \caption{The effective potential  $V_t(z)$ (blue solid lines) and the zero mode $t_0(z)$ (red dashed lines) of the tensor perturbation for brane model 2. The parameters are set as $k=1$,
             and $b=0.5$ in Fig. (a), $b=3$ in Fig. (b), $b=8$ in Fig. (c).} \label{Three}
    \end{figure}

\begin{figure}[!htb]
    \begin{center}
    \subfigure[~$a=0.2$]{\label{figure tensor2}
        \includegraphics[width=3.6cm]{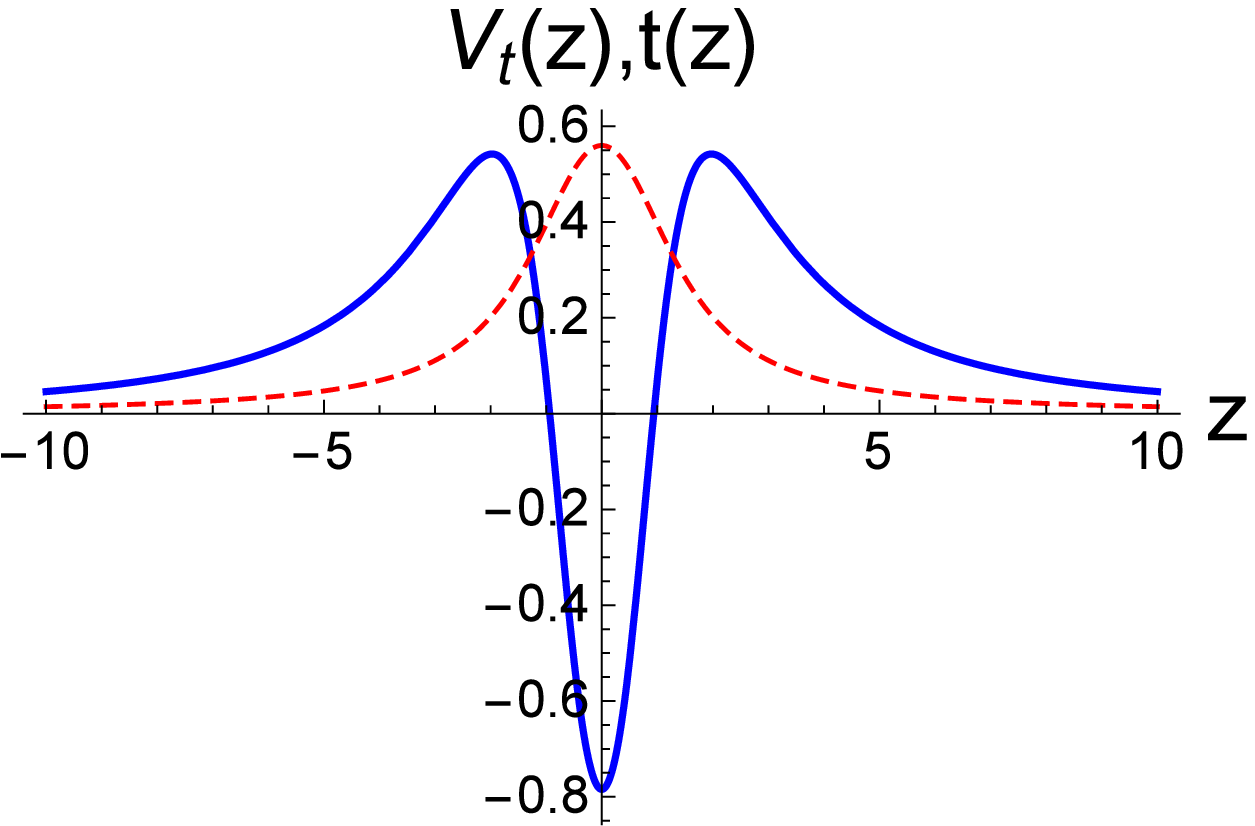}}
    \subfigure[~$b=0.8$]{\label{figure tensor3}
        \includegraphics[width=3.6cm]{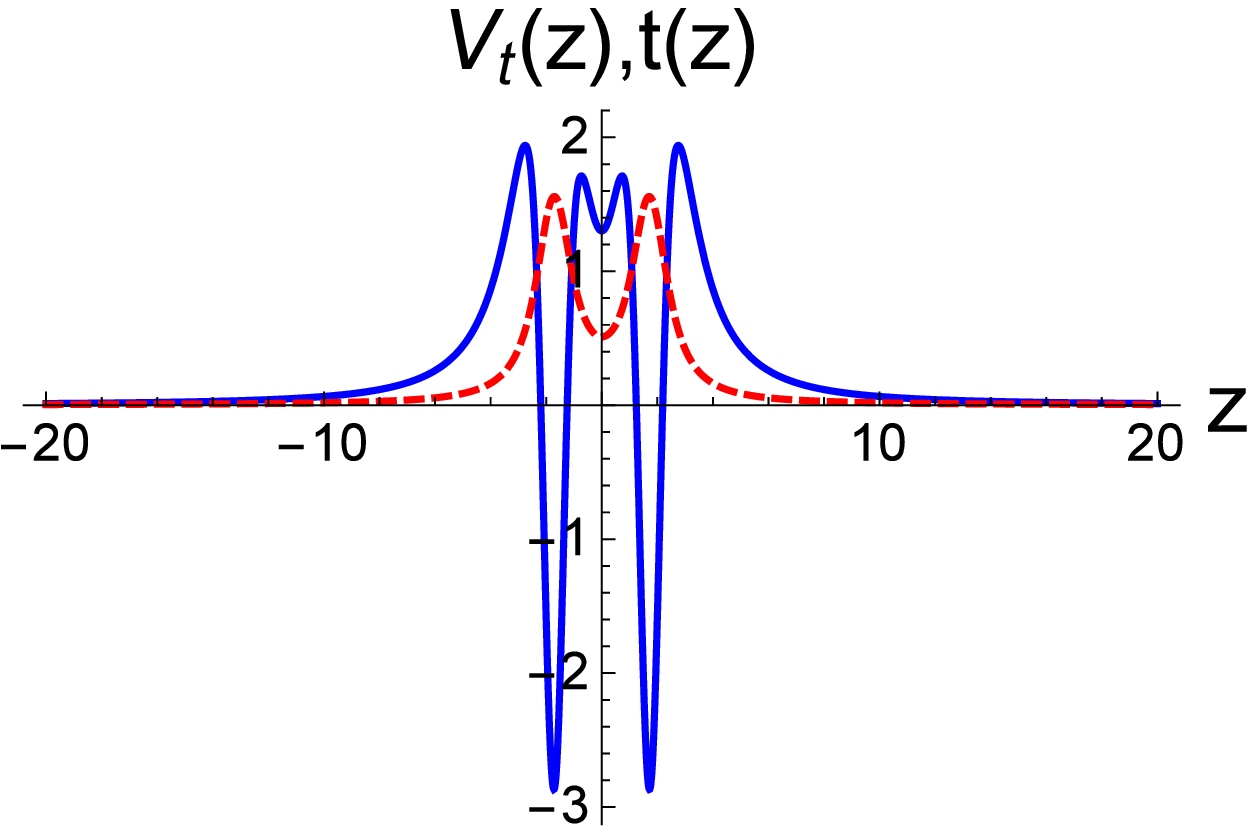}}
    \subfigure[~$b=2.5$]{\label{figure tensor3}
        \includegraphics[width=3.6cm]{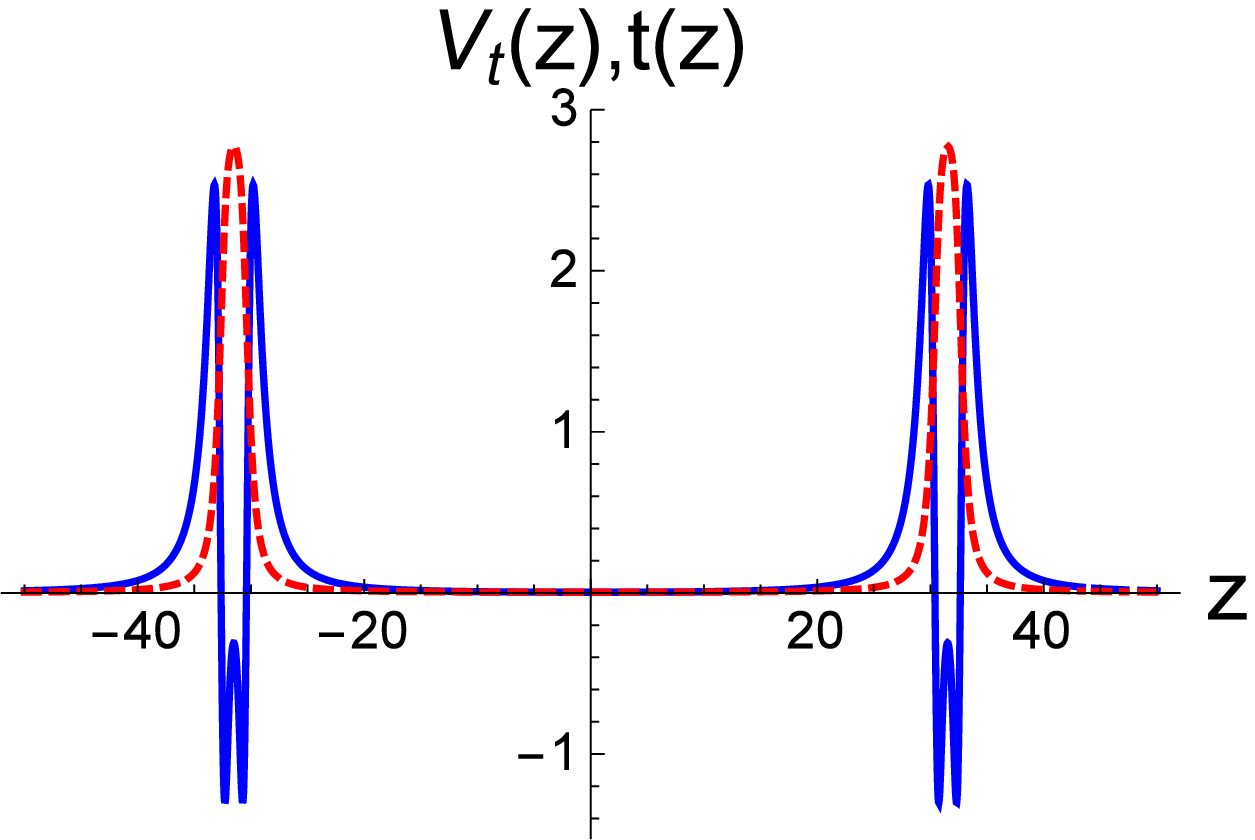}}
    \end{center}
    \caption{The effective potential  $V_t(z)$ (blue solid lines) and the zero mode $t_0(z)$ (red dashed lines) of the tensor perturbation for brane model 3.
    The parameters are set as $k=1$,
             and $a=0.2$ in Fig. (a), $b=0.8$ in Fig. (b), $b=2.5$ in Fig. (c).}\label{Four}
    \end{figure}

In this section, we consider the linear tensor perturbation. Because of the similarity of the field equations between the mimetic gravity and general relativity, it is easy to see that the tensor perturbation is decoupled from the vector and scalar perturbations.  For the tensor perturbation, the perturbed metric is given by
    \begin{eqnarray}
        \label{tensor metric}
        \tilde{g}_{MN}=a(y)^2(\eta_{\mu\nu}+h_{\mu\nu})dx^{\mu}dx^{\nu}+dy^2,
    \end{eqnarray}
where $h_{\mu\nu}=h_{\mu\nu}(x^{\mu},y)$ depends on all the coordinates and satisfies the transverse-traceless (TT) condition $\eta^{\mu\nu}\partial_\mu h_{\lambda\nu}=0$ and $\eta^{\mu\nu}h_{\mu\nu}=0$.
The perturbation of Eq. (\ref{var eom1}) gives
    \begin{eqnarray}
        \frac{1}{2\kappa^2}\delta G_{MN}+\delta\left(\lambda \partial_M \phi \partial_N \phi-\frac{1}{2}L_{\phi}g_{MN}\right)=0.
    \end{eqnarray}
Using this TT condition, the perturbation of the $\mu\nu$ components of the Einstein tensor $\delta G_{\mu\nu}$ reads
    \begin{eqnarray}
    \label{perturbation Gmn}
        \delta G_{\mu\nu}\!\!\!\!&=\!\!\!\!&-\frac{1}{2}\Box^{(4)}h_{\mu\nu}+(3a'^2+3aa'')h_{\mu\nu}\nonumber\\
        \!\!&&-2aa'h'_{\mu\nu}-\frac{1}{2}a^2 h''_{\mu\nu},
    \end{eqnarray}
where the four-dimensional d'Alembertian is defined as $\Box^{(4)}\equiv\eta_{\mu\nu}\partial_{\mu}\partial_{\nu}$.
Using Eqs. (\ref{eom21}) and (\ref{perturbation Gmn}), the above equation reads
    \begin{eqnarray}
        -\frac{1}{2}\Box^{(4)}h_{\mu\nu}
        -2aa'h'_{\mu\nu}-\frac{1}{2}a^2 h''_{\mu\nu}=0.
    \end{eqnarray}
After redefining the extra dimension coordinate $ dz=\frac{1}{a(z)}dy $ and the pertubation
$h_{\mu\nu}=a(z)^{-\frac{3}{2}}\tilde{h}_{\mu\nu}$, we get the equation of $\tilde{h}_{\mu\nu}$:
    \begin{eqnarray}
        \Box^{(4)}\tilde{h}_{\mu\nu}+\partial^2_{z}\tilde{h}_{\mu\nu}
        -\frac{\partial^2_{z}a^{\frac{3}{2}}}{a^{\frac{3}{2}}}\tilde{h}_{\mu\nu}=0.
    \end{eqnarray}
Considering the decomposition $\tilde{h}_{\mu\nu}=\epsilon_{\mu\nu}(x^\gamma) \text{e}^{ip_{\lambda}x^{\lambda}}t(z)$, where the polarization tensor $\epsilon_{\mu\nu}$ satisfies the TT condition $\eta^{\mu\nu}\partial_\mu \epsilon_{\lambda\nu}=0$ and $\eta^{\mu\nu}\epsilon_{\mu\nu}=0$, we obtain the Schr\"{o}dinger-like equation for $t(z)$:
    \begin{eqnarray}
        \label{eq tensor}
        -\partial^2_{z}t(z)+V_t(z)t(z)=m_t^2 t(z),
    \end{eqnarray}
with the potential $V_t(z)$ given by
    \begin{eqnarray}
        V_t(z)=\frac{\partial^2_{z}a^{\frac{3}{2}}}{a^{\frac{3}{2}}}.
    \end{eqnarray}
Now we can see that the equation of the tensor perturbation in mimetic gravity is the same as that in general relativity. Nevertheless, the mimetic scalar field generates more types of thick brane, which could lead to new type of potential of the tensor perturbation. We present the potentials of the tensor perturbations for three models in Figs. \ref{tensorThree} - \ref{Four}, respectively. {In model 1, the potential is a volcano-like potential. As the parameter $n$ increases, the potential well become narrower and deeper.} In model 2, as the parameter $b$ increases, the single brane splits into three sub-branes, and the volcano-like potential changes to a tri-well potential, and at last splits into three independent volcano-like potentials. In model 3,
as the parameter $b$ increases, the single potential well splits into a double-well, and then becomes two volcano-like potentials with inner structure. For both cases, the distance of the those wells increases with $b$.

The zero mode of the tensor perturbation is
    \begin{eqnarray}
        t_0 (z)\propto a^{\frac{3}{2}}(z).
    \end{eqnarray}
It is easy to verify that the zero modes for the above brane models are square-integrable  and hence are localized around the brane. Thus the four-dimensional Newtonian potential can be realized on the brane. Also Eq. (\ref{eq tensor}) can be factorized as
    \begin{eqnarray}
        \label{eq tensor2}
        \left(-\partial_{z}+\frac{\partial_{z}a^{\frac{3}{2}}}{a^{\frac{3}{2}}}\right)
        \left(\partial_{z}+\frac{\partial_{z}a^{\frac{3}{2}}}{a^{\frac{3}{2}}}\right)t(z)
        =m_t^2 t(z).
    \end{eqnarray}
It is clear that there is no tensor tachyon mode, thus the brane is stable against the tensor perturbation.

\section{Scalar perturbation} \label{Sec4}
In this section, we study the scalar pertubation.
The perturbed metric is
    \begin{eqnarray}
        \label{brane metric}
       ds^2=a^{2}(z)\left[(1+2\psi)\eta_{\mu\nu}dx^{\mu}dx^{\nu}+(1+2\Phi)dz^2\right].
    \end{eqnarray}
From Eq. (\ref{var eom1}) we have
    \begin{eqnarray}
        \label{eom5}
        R_{MN}+2\lambda \partial_M\phi\partial_N\phi-\frac{2}{3}g_{MN}(\lambda U+V)=0.
    \end{eqnarray}
The perturbation of Eq. (\ref{eom5}) reads
    \begin{eqnarray}
        \delta R_{\mu\nu}-\frac{4}{3}(\partial_z a)^{2}\psi(\lambda U+V)\eta_{\mu\nu} \nonumber\\
        -\frac{2}{3}a^2\eta_{\mu\nu}
        \left(\lambda \frac{\partial U}{\partial \phi}\delta\phi+\frac{\partial V}{\partial \phi}\delta\phi \right)=0,\\
        \delta R_{\mu5}+2\lambda\partial_z \phi\partial_\mu \delta \phi=0,
        \label{scalar pertb mu5}\\
        \delta R_{55}+4\lambda\partial_z \phi\delta\partial_z \phi-\frac{2}{3}a^2\left(\lambda \frac{\partial U}{\partial \phi}\delta\phi+\frac{\partial V}{\partial \phi}\delta\phi \right) \nonumber\\
        -\frac{4}{3}a^2(\lambda U+V)\Phi=0,
    \end{eqnarray}
where the components of $\delta R_{MN}$ are given by
    \begin{eqnarray}
        \label{ptb1}
        \delta R_{\mu\nu}\!\!\!&=&\!\!\!-2\partial_{\mu}\partial_{\nu}\psi
        -\partial_{\mu}\partial_{\nu}\Phi-\eta_{\mu\nu}\Box^{(4)}\psi-\eta_{\mu\nu}\partial^2_z \psi,\nonumber\\
        &&\!\!\!\!\!\!+\left(\frac{4(\partial_z a)^2}{a^2}+\frac{2\partial^2_z a}{a}\right)(\Phi-\psi)\eta_{\mu\nu}\nonumber\\
        &&+\frac{\partial_z a}{a}(\partial_z \phi-7\partial_z \psi)\eta_{\mu\nu},\\
        \label{ptb2}
        \delta R_{\mu 5}\!\!\!&=&\!\!\! \partial_\mu \left(\frac{3\partial_z a}{a}\Phi-3\partial_z \psi\right),\\
        \label{ptb3}
        \delta R_{55}\!\!\!&=&\!\!\!-\Box^{(4)}\Phi-4\partial^2_z \psi+\frac{4\partial_z a}{a}(\partial_z \phi-\partial_z \psi).
    \end{eqnarray}

On the other hand, the perturbation of Eq. (\ref{var eom3}) gives
    \begin{eqnarray}
        \label{ptb4}
        \frac{2}{a^2}\partial_z \phi \partial_z \delta \phi-\frac{2}{a^2}(\partial_z \phi)^2\Phi=\frac{\partial U}{\partial \phi}\delta\phi,
    \end{eqnarray}
which follows that
    \begin{eqnarray}
        \label{Phi dt phi}
        \Phi=\frac{\partial_z \delta\phi}{\partial_z \phi}-\frac{a^2}{2(\partial_z \phi)^2}\frac{\partial U}{\partial \phi}\delta\phi.
    \end{eqnarray}
From the off-diagonal part of Eq. (\ref{ptb1}) we get the simple relation between the scalar modes  $\Phi$ and $\psi$ in the perturbation of the metric:
    \begin{eqnarray}
        \label{ptb5}
        \Phi=-2\psi.
    \end{eqnarray}
Substituting Eqs. (\ref{Phi dt phi}) and (\ref{ptb5}) into Eq. (\ref{scalar pertb mu5}) and integrating with respect to the four-dimensional coordinates $x^{\mu}$, we get the master equation of the scalar perturbation $\delta\phi$
    \begin{eqnarray}
       && \!\!\!\!\!\!\!\!\!\!\!\!\!\!\!\!\!\!\!\!\!\!\!\!\label{scalar master}
    -\frac{3}{2}\partial^2_z \delta\phi
         +\frac{3}{4}\left(\frac{a^2}{\partial_z \phi}\frac{\partial U}{\partial \phi}+\frac{2\partial^2_z \phi}{\partial_z \phi}-\frac{4\partial_z a}{a}\right)\partial_z \delta \phi    \nonumber\\
        && \!\!\!\!\!\!\!\!\!\!\!\!\!\!\!\!\!\!\!\!\!\!\!\!\!\!\!\!\!\!\!\!\!\!+\!\left[\frac{3a\partial_z a}{\partial_z \phi}\frac{\partial U}{\partial \phi}\!\!+\!\!2\lambda(\partial_z \phi)^2\!\!+\!\!\frac{3}{4}a^2\left(\frac{\partial^2 U}{\partial \phi^2}\!-\!2\frac{\partial U}{\partial \phi}\frac{\partial^2_z \phi}{(\partial_z \phi)^2}\right)\right]\!\delta\phi\!\!=\!\!0.
    \end{eqnarray}
To simplify this equation, we have to use the background equations (\ref{var eom1})-(\ref{var eom3}) in the coordinate system ($x^{\mu},z$),
    \begin{eqnarray}
        \label{eom1}
       \!\!\!\!\!\!\!\!\!\!\!\!\frac{3a''}{a^3}=-V(\phi), \\
       \label{eom2}
       \!\!\!\!\!\!\!\!\!\!\!\!\frac{6a'^2}{a^2}+a^2 V(\phi)+2a^2\lambda U(\phi)=0, \\
       \label{eom3}
       a^2(\lambda U'(\phi)+\frac{\partial V}{\partial \phi})+6\lambda\partial_z \phi\frac{a'}{a}+2\lambda'\partial_z \phi\nonumber\\
       +2\lambda\partial^2_z \phi=0,\\
       \label{eom4}
       \!\!\!\!\!\!\!\!\!\!\!\!\frac{1}{a^2}(\partial_z \phi)^2=U(\phi),
    \end{eqnarray}
 and redefine $\delta\phi(x^{\mu},z)=\frac{(\partial_z \phi)^{\frac{3}{2}}}{a^2}s(z)\overline{\delta\phi}(x^\mu)$. Then Eq. (\ref{scalar master}) turns to
     \begin{eqnarray}
        \label{scalar pertb}
        -\partial^2_z s(z)+V_s(z)s(z)=0,
    \end{eqnarray}
 with the effective potential $V_s(z)$ given by
    \begin{eqnarray}
        V_s(z)=\frac{2(\partial_z a)^2-a\partial^2_z a}{a^2}
        +\frac{-(\partial^2_z \phi)^2+2\partial_z \phi \partial^{3}_z\phi}{4(\partial_z\phi)^2}.
    \end{eqnarray}
The corresponding scalar perturbation mode in the metric is given by Eqs. (\ref{Phi dt phi}) and (\ref{ptb5}).

Note that there is no term of the form $\Box^{4} \delta\phi$ in Eq. (\ref{scalar master}), and hence there is no term of the form $m_s^2 s(z)$ in  Eq. (\ref{scalar pertb}), which is consistent with the cosmological scalar perturbation in mimetic gravity \cite{Chamseddine:2014vna}.  This implies that the scalar perturbations do not propagate on the brane. Thus there is no tachyon scalar mode, and the brane is stable under the linear scalar perturbations.

\begin{figure}[!htb]
    \begin{center}
    \subfigure[~$n=1$]{\label{figure scalar11}
        \includegraphics[width=3.6cm]{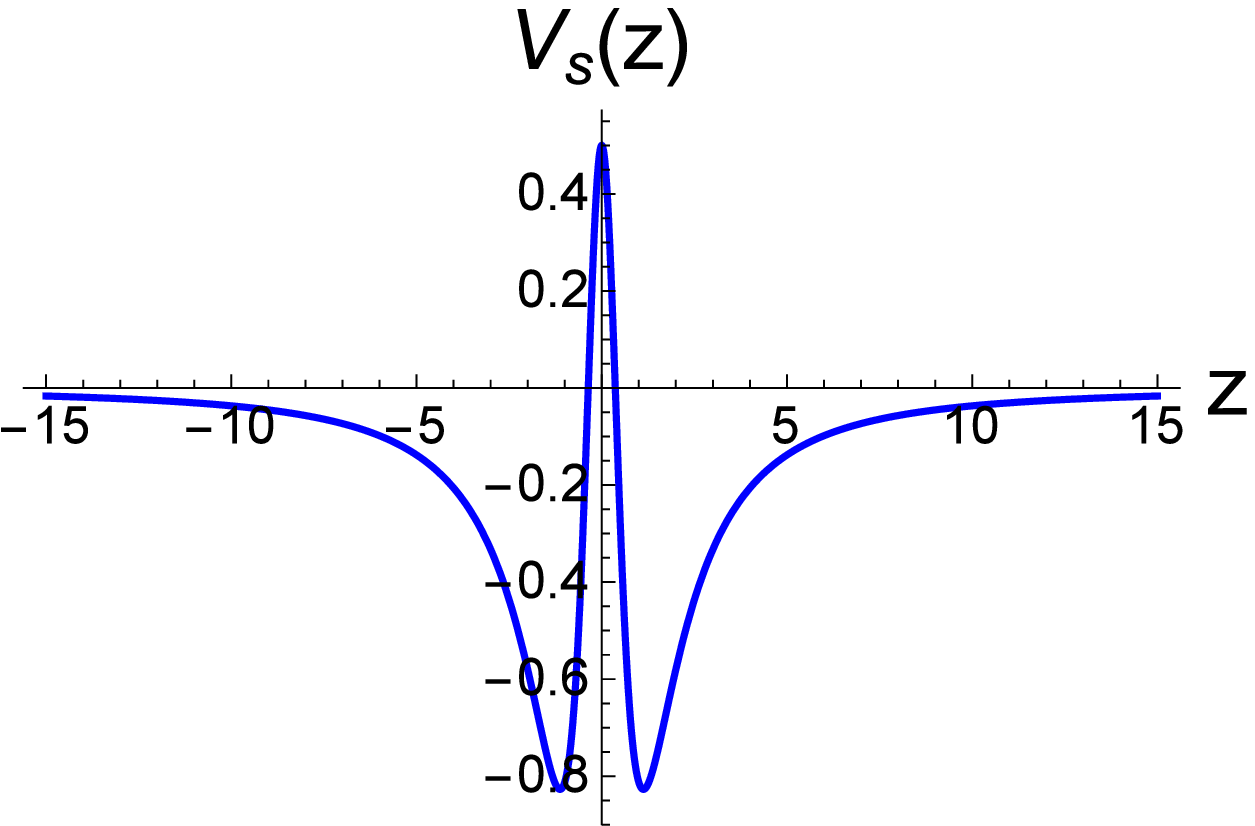}}
    \subfigure[~$n=3$]{\label{figure scalar12}
        \includegraphics[width=3.6cm]{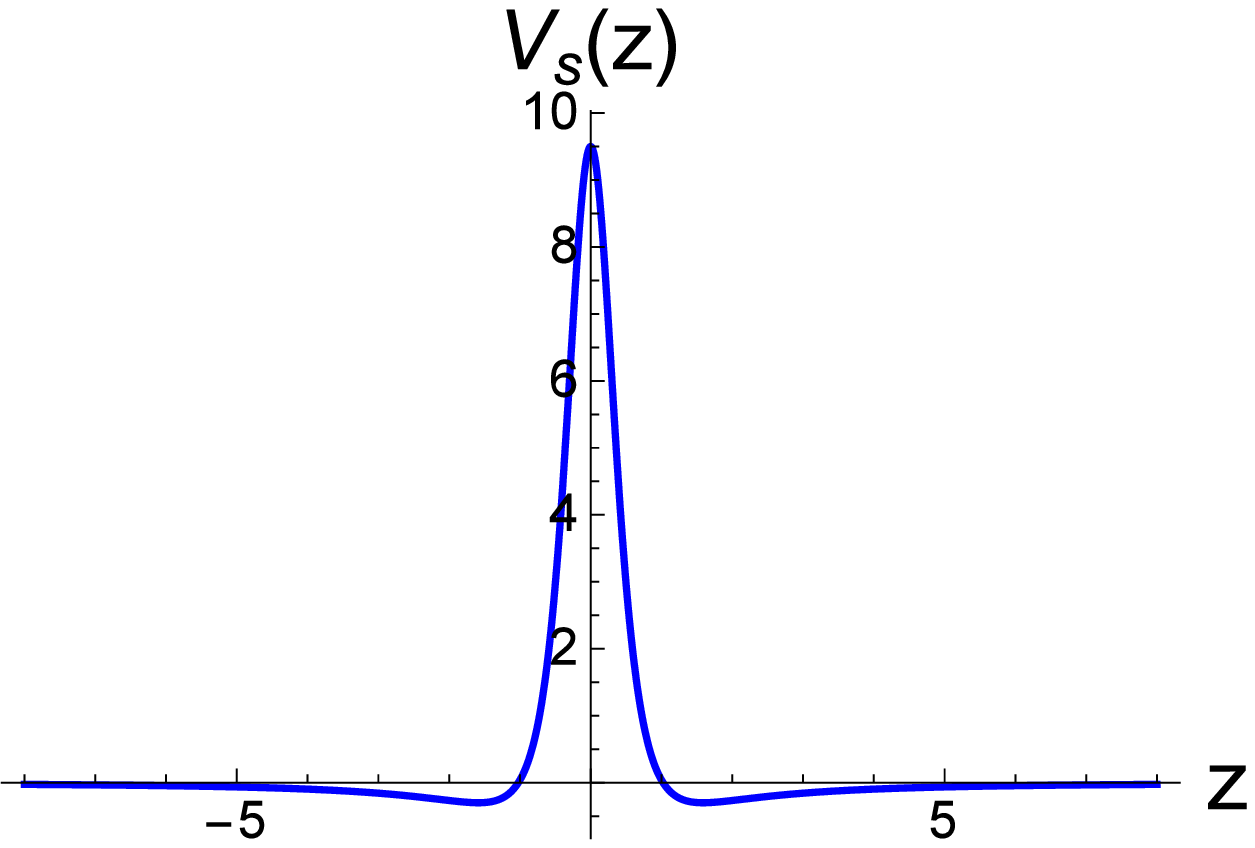}}
    \subfigure[~$n=5$]{\label{figure scalar13}
        \includegraphics[width=3.6cm]{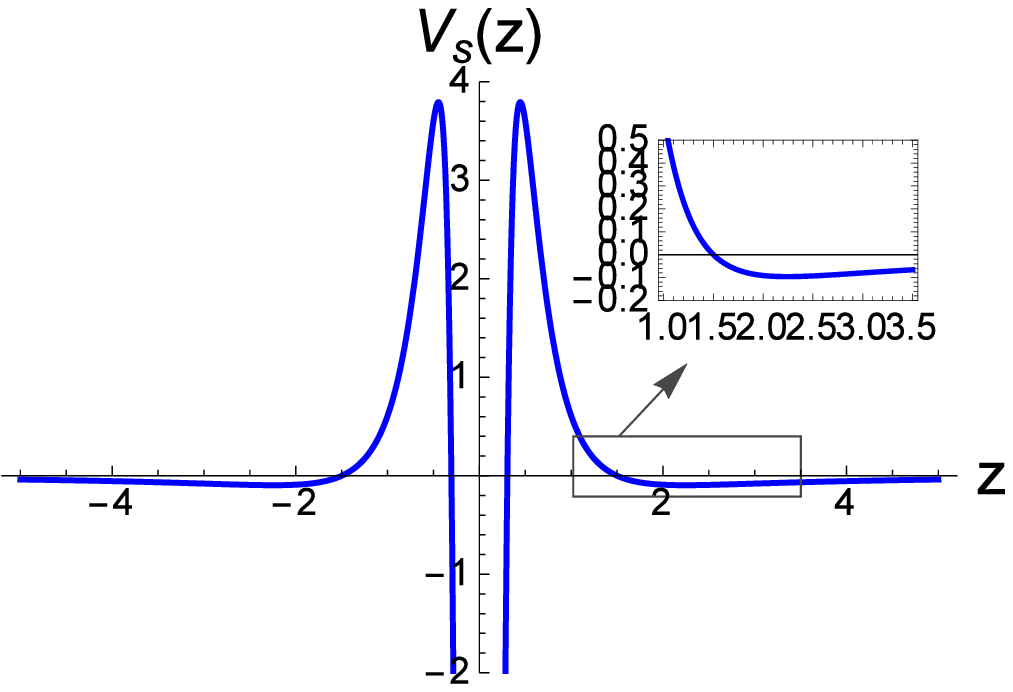}}
    \end{center}
    \caption{The effective potential  $V_s(z)$ for model 1. The parameters are set as $k=1$, $v=1$
             and $n=1$ in Fig. (a), $n=3$ in Fig. (b), $n=5$ in Fig. (c).} \label{figVs_model1}
    \end{figure}

\begin{figure}[!htb]
    \begin{center}
    \subfigure[~$b=0.5$]{\label{figure tensor2}
        \includegraphics[width=3.6cm]{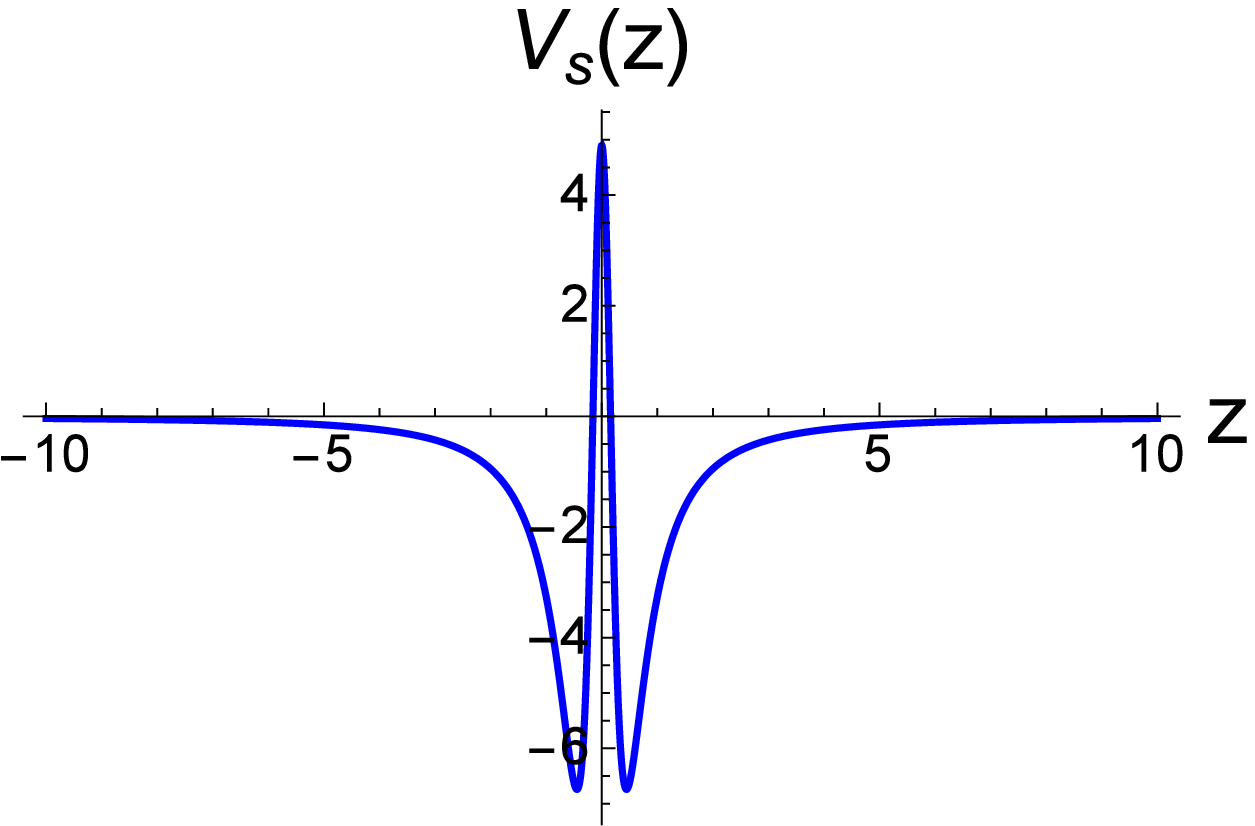}}
    \subfigure[~$b=3$]{\label{figure tensor3}
        \includegraphics[width=3.6cm]{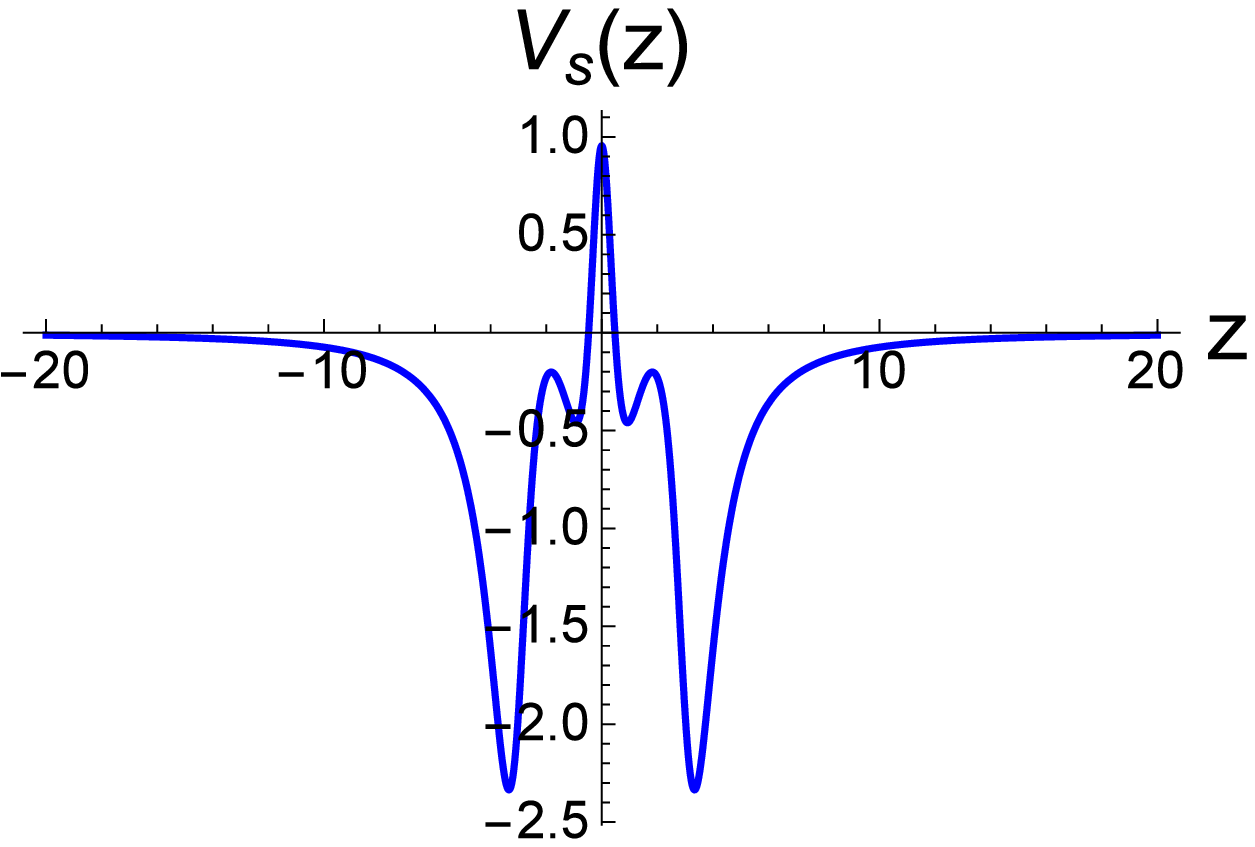}}
    \subfigure[~$b=8$]{\label{figure tensor3}
        \includegraphics[width=3.6cm]{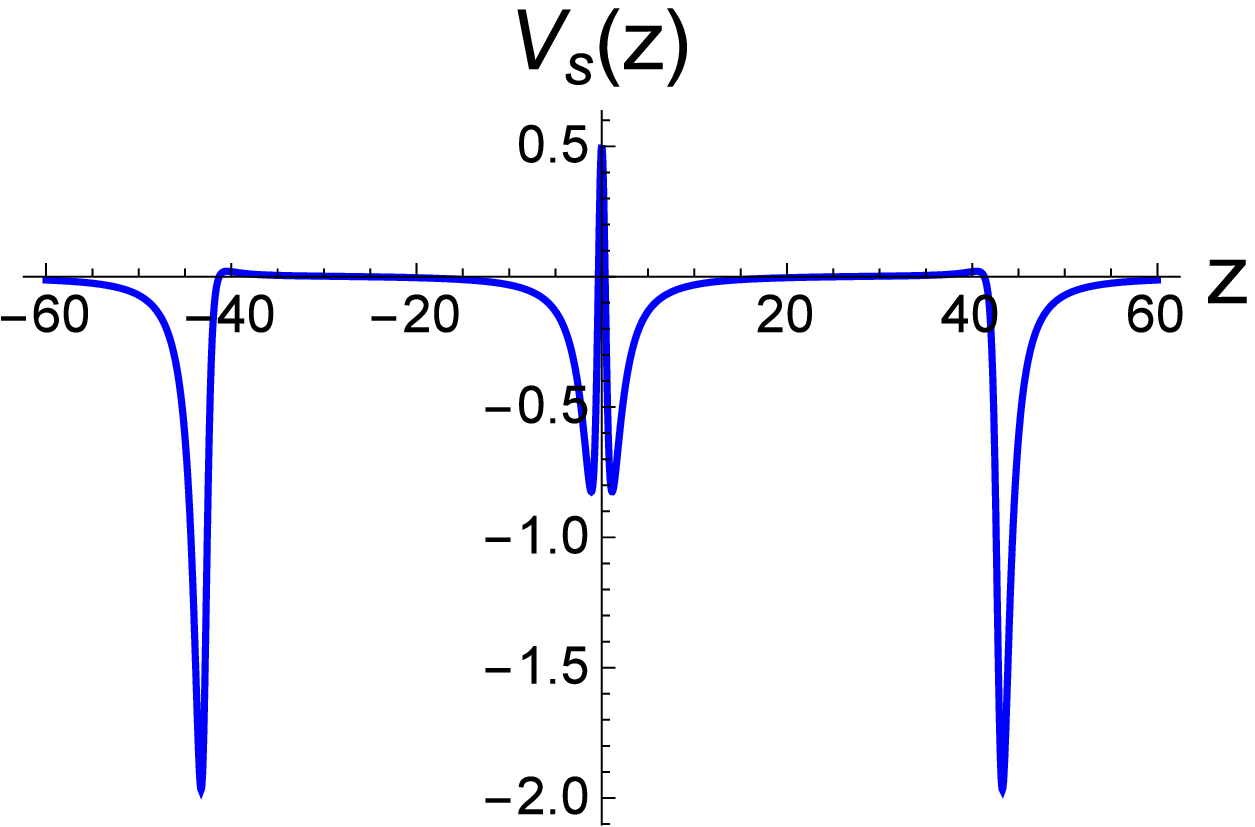}}
    \end{center}
    \caption{The effective potential  $V_s(z)$ for model 2. The parameters are set as $k=1$, $v=1$
             and $b=0.5$ in Fig. (a), $b=3$ in Fig. (b), $b=8$ in Fig. (c).} \label{figVs_model2}
    \end{figure}

\begin{figure}[!htb]
    \begin{center}
    \subfigure[~$a=0.2$]{\label{figure tensor2}
        \includegraphics[width=3.6cm]{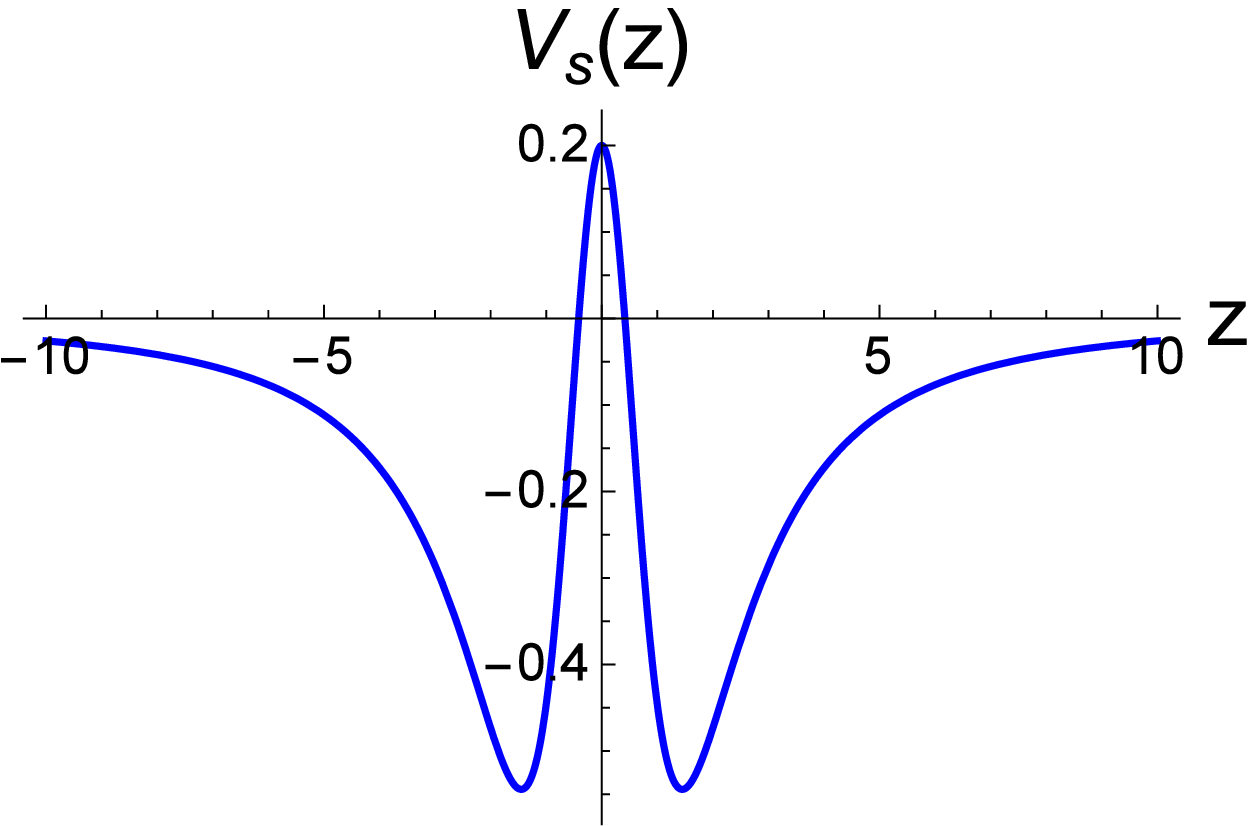}}
    \subfigure[~$b=0.8$]{\label{figure tensor3}
        \includegraphics[width=3.6cm]{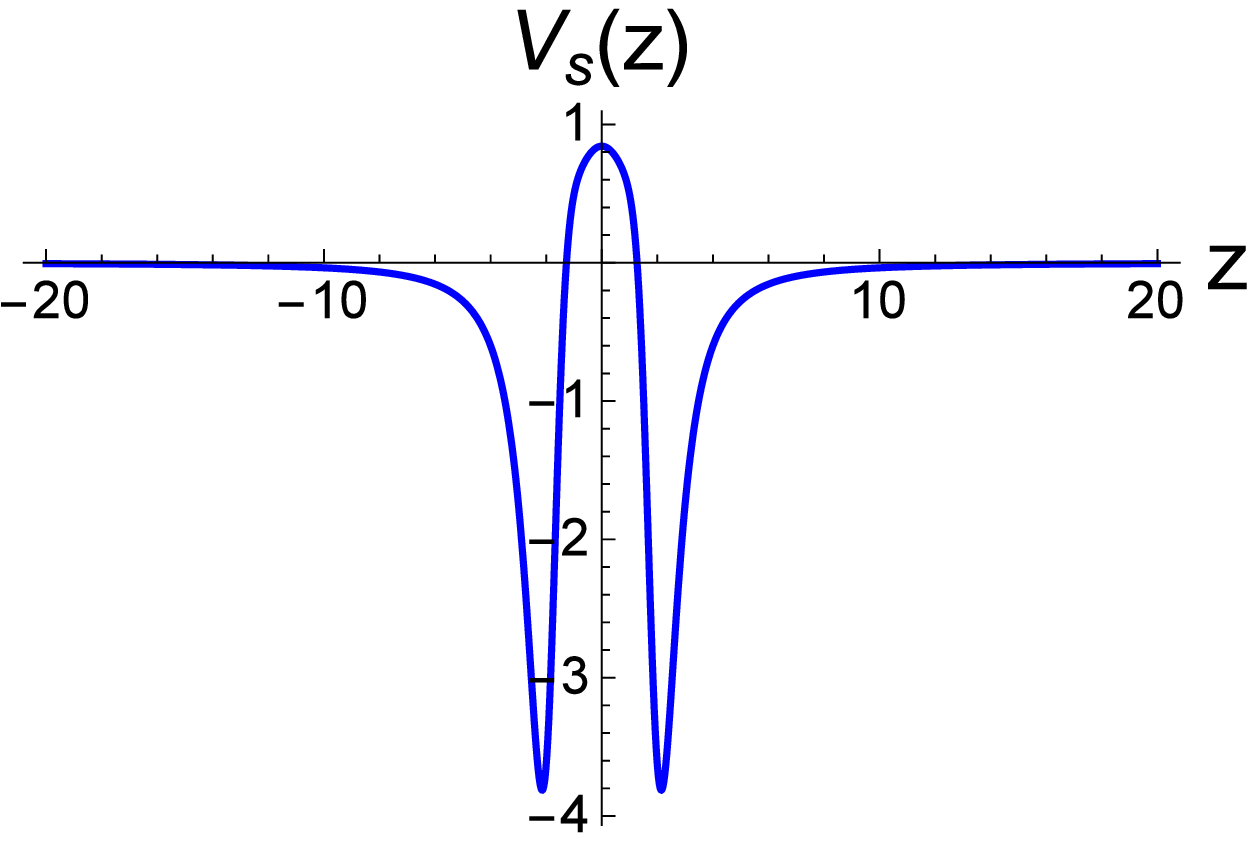}}
    \subfigure[~$b=2.5$]{\label{figure tensor3}
        \includegraphics[width=3.6cm]{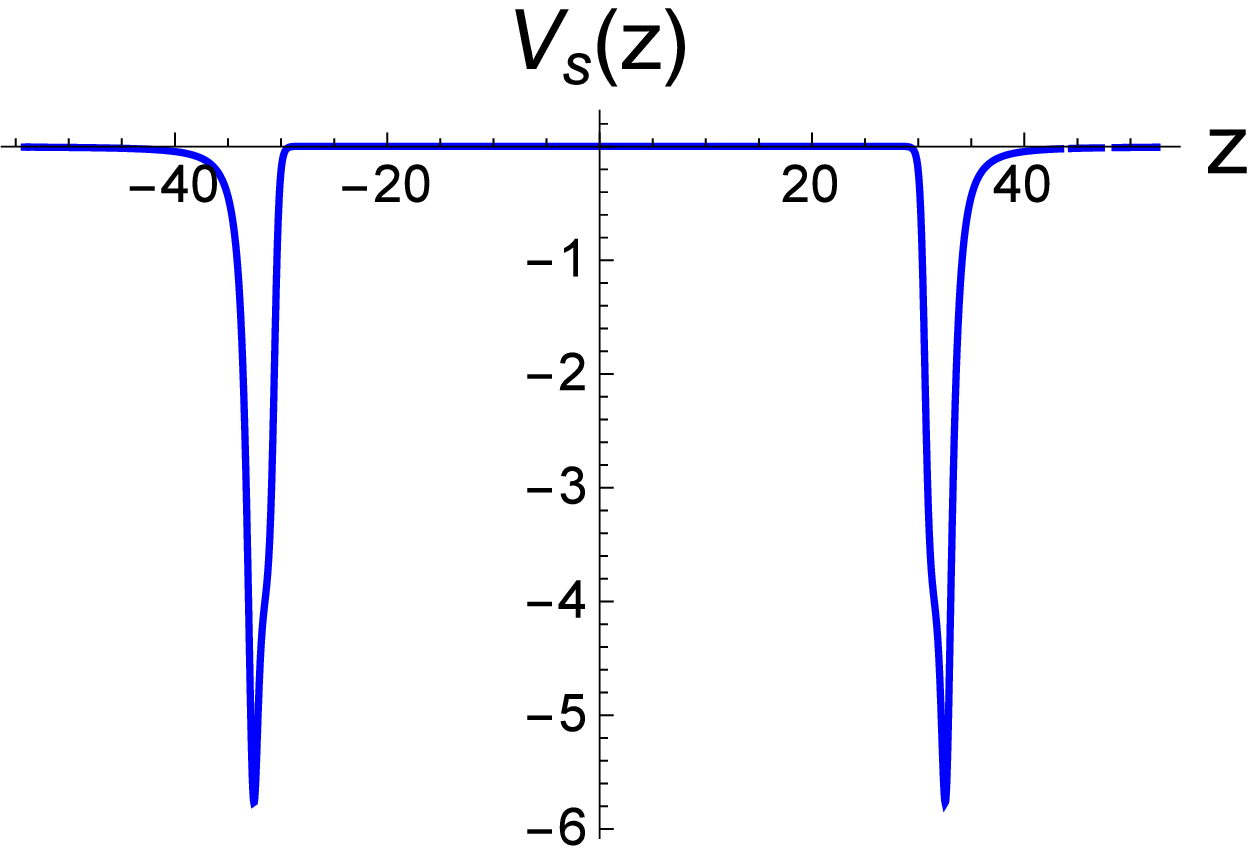}}
    \end{center}
    \caption{The effective potential  $V_s(z)$ of the model 3. The parameters are set as $k=1$, $v=1$, $n=1$,
             and $a=0.2$ in Fig. (a), $b=0.8$ in Fig. (b), $b=2.5$ in Fig. (c).}\label{figVs_model3}
    \end{figure}

The effective potential  $V_s(z)$ for the three models
are shown in Figs.~\ref{figVs_model1} - \ref{figVs_model3}, respectively. From these figures, it can be seen that, {for model 1, there are two wells for the parameter $n=1$ and $n=3$, while when $n=5$ there are three, and the potential is divergent at the origin; for model 2, the potential turns from a double-well into four sub-wells as the parameter $b$ increases; for model 3, the potential remains a double-well as the parameter $b$ increases.} Furthermore, for all the three brane models, the potential approaches $0^-$ at infinity, hence the scalar perturbations are not localized on the brane and would not lead to the ``fifth force''.

\section{Conclusion} \label{SecConclusion}
In this work, we investigated three kinds of thick branes generated by the mimetic scalar field, which represents the isolated conformal degree of freedom. Since we are free to choose arbitrary potentials $V(\phi)$ and $U(\phi)$, it is possible to construct abundant kinds of thick brane models in mimetic gravity.
In the first brane model, we get a single brane with a double-kink scalar field.
In the last two brane models, the branes split into many sub-branes as the parameter $b$ increases, and the potentials $V_t(z)$ and $V_s(z)$ of the extra parts $t(z)$ and $s(z)$ of the tensor and scalar perturbations also split into multi-wells. We also showed that the branes are stable under the tensor perturbations and the Newtonian potentials can be realized on the branes.
The scalar perturbations do not propagate on the brane, which is quite different from other brane models.
By analyzing the potential $V_s(z)$ we conclude that the scalar perturbations $s(z)$ for the three models  are not localized on the brane.

Furthermore, models 2 and 3 can be extended into the case of brane array. The inner structure of the brane may lead to new phenomenon in the resonance of the tensor perturbation and the localization of matter fields. We will consider this issue in the future work.

\section*{Acknowledgement}

This work was supported by the National Natural Science Foundation of China (Grants Nos. 11522541, 11375075, and 11605127) and the Fundamental Research Funds for the Central Universities (Grants No. lzujbky-2016-k04 and No. lzujbky-2017-it68). Yuan Zhong was also supported by China Postdoctoral Science Foundation (Grant No. 2016M592770).


\end{document}